\documentclass[twocolumn]{article}
\usepackage[margin=0.7in]{geometry}

\usepackage{graphicx} 
\usepackage{subfig}
\usepackage{mathrsfs}
\usepackage{rsfso}
\usepackage{amsfonts} 
\usepackage{amsmath}
\usepackage{booktabs} 
\usepackage{multirow}
\usepackage{dsfont}
\usepackage{todonotes}
\usepackage{pdfpages}
\usepackage[linesnumbered,ruled,vlined]{algorithm2e}
\usepackage{hyperref}
\usepackage{amsmath}
\usepackage{acro}
\usepackage{bbm}
\usepackage{color}
\usepackage{authblk}
\date{}
\usepackage[numbers]{natbib}
\DeclareAcronym{ATV}{
  short = ATV,
  long  = Average Total Variation,
}
\definecolor{headercolor}{HTML}{D9EAD3}

\begin{document}

\title{Modeling Musical Genre Trajectories through Pathlet Learning}

\author[1]{Lilian Marey}
\author[2]{Charlotte Laclau}
\author[3]{Bruno Sguerra}
\author[4]{Tiphaine Viard}
\author[5]{Manuel Moussallam}

\affil[1]{LTCI, Télécom Paris - Deezer Research\\
\texttt{mareylilian@gmail.com}}
\affil[2]{LTCI, Télécom Paris\\
\texttt{charlotte.laclau@telecom-paris.fr}}
\affil[3]{Deezer Research\\
\texttt{bmassonisguerra@deezer.com}}
\affil[4]{i3, Télécom Paris\\
\texttt{tiphaine.viard@telecom-paris.fr}}
\affil[5]{Deezer Research\\
\texttt{manuel.moussallam@deezer.com}}
% \author{Lilian Marey}
% \email{mareylilian@gmail.com}
% \orcid{0000-0002-8518-1431}
% \affiliation{%
%   \institution{LTCI, Télécom Paris}
%   \city{Palaiseau}
%   \country{France}
% }
% \affiliation{%
%   \institution{Deezer Research}
%   \city{Paris}
%   \country{France}
% }

% \author{Charlotte Laclau}
% \email{charlotte.laclau@telecom-paris.fr}
% \orcid{0000-0002-8518-1431}
% \affiliation{%
%   \institution{LTCI, Télécom Paris}
%   \city{Palaiseau}
%   \country{France}
% }

% \author{Bruno Sguerra}
% \email{bmassonisguerra@deezer.com}
% \orcid{0000-0003-1158-9095}
% \affiliation{%
%   \institution{Deezer Research}
%   \city{Paris}
%   \country{France}
% }

% \author{Tiphaine Viard}
% \email{tiphaine.viard@telecom-paris.fr}
% \orcid{0000-0002-5969-5439}
% \affiliation{%
%   \institution{i3, Télécom Paris}
%   \city{Palaiseau}
%   \country{France}
% }

% \author{Manuel Moussallam}
% \email{manuel.moussallam@deezer.com}
% \orcid{0000-0003-0886-5423}
% \affiliation{%
%   \institution{Deezer Research}
%   \city{Paris}
%   \country{France}
% }

% \renewcommand{\shortauthors}{Marey et al.}

% \begin{CCSXML}
% <ccs2012>
%    <concept>
%        <concept_id>10003120.10003121.10003122.10003332</concept_id>
%        <concept_desc>Human-centered computing~User models</concept_desc>
%        <concept_significance>500</concept_significance>
%        </concept>
%  </ccs2012>
% \end{CCSXML}

% \ccsdesc[500]{Human-centered computing~User models}

% \keywords{user modeling, multimedia streaming, user trajectories, dictionary learning, pathlet learning}

\maketitle

\begin{abstract}
The increasing availability of user data on music streaming platforms opens up new possibilities for analyzing music consumption. 
However, understanding the evolution of user preferences remains a complex challenge, particularly as their musical tastes change over time.
This paper uses the dictionary learning paradigm to model user trajectories across different musical genres. 
We define a new framework that captures recurring patterns in genre trajectories, called \textit{pathlets}, enabling the creation of comprehensible trajectory embeddings. 
We show that pathlet learning reveals relevant listening patterns that can be analyzed both qualitatively and quantitatively. 
This work improves our understanding of users' interactions with music and opens up avenues of research into user behavior and fostering diversity in recommender systems.
A dataset of 2000 user histories tagged by genre over 17 months, supplied by Deezer (a leading music streaming company), is also released with the code. 
\end{abstract}

\section{Introduction}
\label{intro}

\begin{figure*}[ht]
\centering
\includegraphics[width=00.9\linewidth]{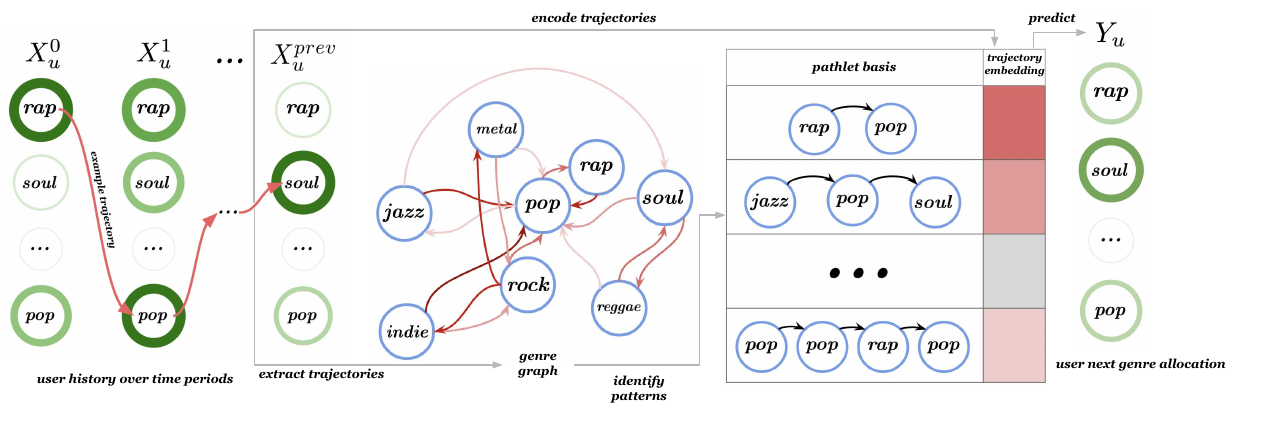}
\caption{Our methodology for users preferences modeling. 
Music genre trajectories are extracted from user listening histories and processed using the pathlet learning algorithm. This highlights recurring patterns in user behavior, which are then used to generate explicit trajectory embeddings that predict future genre allocation.}
\label{fig1}
\end{figure*}

The advent of music streaming platforms has transformed the way people engage with music, providing both industry and research scientists with access to extensive datasets that capture the intricate interactions between users and vast musical catalogs. 
This shift thus opens up new possibilities for analyzing listening behaviors and musical tastes over time \cite{cura2022uplifting, moore2013taste, petridis2022tastepaths, seaver2022computing, villermet2021follow}.

On the one hand, it is of particular interest to sociology, as music plays a crucial role in social life, shaping identities, group affiliations, and cultural dynamics. 
Indeed, a whole literature demonstrates that listening practices are not merely individual preferences but are embedded in broader social structures, influenced by factors such as class, age, gender, and technological mediation \cite{bourdieu2018distinction, denora2000music, hesmondhalgh2013music}. 
Understanding how people engage with music—what they listen to, how they discover new tracks, and how their tastes evolve—offers valuable insights into social stratification, cultural capital, and the ways in which digital platforms mediate contemporary cultural consumption \cite{mccourt2016music, riom2020discovering}.

Moreover, this topic is highly relevant to the music industry, particularly in the development of recommendation systems. 
These systems play a crucial role in shaping users' listening experiences by suggesting songs, albums, and artists based on their past behaviors, preferences, and broader consumption patterns \cite{schedl2018current}. 

However, the increasing reliance on data-driven recommendation systems has also led to a race for performance optimization, driven by the quantification of user engagement and algorithmic efficiency. 
This pursuit raises new challenges, particularly with the widespread use of black-box models, whose decision-making processes remain opaque \cite{bucher2016neither}. As platforms prioritize metrics such as click-through rates and listening duration, there is a risk of reinforcing homogenized consumption patterns, limiting musical diversity and amplifying biases embedded in training data \cite{lesota2022traces, matrosova2024recommender}. 
These concerns call for a critical examination of the sociotechnical dynamics at play, questioning the long-term implications of algorithmic mediation in cultural consumption.

In response to these issues, user studies have emerged as a crucial approach to understanding how listeners and providers perceive and experience algorithmic recommendations \cite{ungruh2024putting, vrijenhoek2024diversity, smith2024recommend, ekstrand2014user}. Research in this area explores users' trust in recommendation systems, their sense of control over music discovery, and the perceived trade-offs between personalization and diversity. 
These studies highlight the importance of designing recommendation systems that align not only with engagement metrics but also with user satisfaction, autonomy, and the richness of musical exploration.

This observation underscores the need for designing explainable methods that move beyond black-box models, offering greater transparency and interpretability to users.
Instead of prioritizing optimizing engagement metrics, research is increasingly exploring techniques such as interpretable machine learning, rule-based recommendations, and hybrid approaches \cite{moore2013taste, afchar2022explainability} that balance accuracy with explainability. 

Developing such approaches is crucial not only for increasing transparency and user trust, but also for addressing ethical concerns related to algorithmic biases, filter bubbles, and the broader impact of automated curation on musical diversity.

In this spirit, our paper aims to model the evolution of musical taste over time, with a particular focus on music genres. 
While grounded in a quantitative approach, our framework offers transparency through the use of a dictionary learning based approach, built to capture the dynamics of genre emergence and disappearance for users. 
More concretely, by representing user listening trajectories through vectors with explicit dimensions, we provide an understandable way to represent musical consumption, which can be useful both for transparency in music recommendation systems and sociology of cultural practices. 

\subsection*{Contributions}
This paper aims to model the evolution of music consumption on streaming platform, while highlighting listening trends through the use of explicit embeddings.
The key contributions are as follows: 
\begin{enumerate}
    \item We propose a methodology for modeling user listening trajectories, focusing on the long-term evolution of music consumption by genre. Building on the dictionary learning paradigm, the method captures recurring patterns in users' preferences over time, which serves transparency purposes.
    \item Through this framework, we analyze the long-term evolution of genre consumption, revealing valuable insights such as inherent mechanisms of interaction between different music genres at play in the formation of musical taste.
    \item We release a proprietary dataset from Deezer (a major music streaming platform), consisting of 2000 user histories labeled by genre over a 17-month period, along with our code and experiments, available at \href{https://github.com/lilianmarey/music_pathlets}{github.com/lilianmarey/music\_pathlets} and \href{https://zenodo.org/records/15341401}{zenodo.org/records/15341401}.
\end{enumerate}

This paper is organized as follows: In Section~\ref{section2}, we present the background and motivation for our study. Section~\ref{section3} introduces our approach to pathlet learning for user trajectories (see Figure~\ref{fig1}), while we outline the experimental protocol in Section~\ref{section4}.
Finally, we present our results in Section~\ref{section5}.

\section{Related works}

For modeling musical user taste, classical approaches have primarily focused on techniques such as collaborative filtering \cite{zhou2008large, hu2008collaborative, takacs2012alternating}, content-based filtering \cite{van2000using}, and hybrid methods \cite{burke2002hybrid}. 
Collaborative filtering relies on user-item interactions or explicit feedback to identify patterns in user preferences \cite{jawaheer2014modeling}, and uses similarity metrics to recommend music based on the past behaviors of similar users. In contrast, content-based filtering suggests music based on the attributes of the content itself, such as genre, artist, or tempo.
In addition, recent advancements in machine learning have introduced deep learning-based recommender systems that model complex relationships between users and music, improving the accuracy of predictions and enabling better handling of large datasets \cite{he2017neural}. 

More recently a key focus is the development of models specifically designed to capture long-term musical preferences. 
Techniques such as recurrent neural networks \cite{medsker2001recurrent},  long short-term memory networks \cite{graves2012long} or transformers \cite{tran2024transformers} have been applied to predict and understand shifts in musical taste based on historical data \cite{devooght2017long, deepak2020music}. However, these models are often considered black-box approaches, as they typically lack transparency in their decision-making processes, making it difficult for users to understand how their preferences are being modeled and predicted.

To counteract this opacity problem, the use of trajectory-based models has proved to be relevant. For instance, as presented in \cite{moore2013taste} and \cite{RomeroB22}, building temporal embeddings through tasks such as playlist continuation enables a dynamic understanding of how musical preferences evolve. 
These approaches allow us to visualize how embeddings—representing users' tastes or musical genres—shift and evolve over time.

One specific application of trajectory modeling is genre transition prediction. Introduced by Sanna et al., the Preference Transition Model \cite{sanna2021next} overcomes the matrix based models \textit{Non-negative Matrix Factorization} \cite{zhang2006learning} and \textit{Poisson Matrix Factorization}, \cite{gopalan2015scalable} providing an explicit music genre transition tool.
However, although the paper reveals valuable insights into the proximity between genres, it only considers interactions at the first-order level, focusing solely on transitions between two periods and interactions between just two genres at a time.
By focusing on trajectory modeling through graphs, our presented approach can align with the domain of graph-based recommendation systems \cite{guo2020survey}. 
These works allow capturing more complex relationships through non-Euclidean data structures, extending beyond simple pairwise interactions. 
Specifically, in \cite{hou2019explainable, wu2023generic}, the authors leverage knowledge graphs and Markov decision processes to enable qualitative analysis of user-item interactions through path-based methods, in a session-based recommendation task. 
However, while these methods provide complementary information to user-item interactions for short-term next-item recommendations, our approach focuses on explaining long-term user preference evolution, offering broader listening patterns.

Ultimately, the algorithmic method presented in our paper draws on the Dictionary Learning (DL) paradigm \cite{mairal2009online, wohlberg2014efficient, grosse2012shift, marey2024modeling}. 
Specifically, these ideas are applied in a graph context, namely through Pathlet Learning (PL). 
While the literature have demonstrated the effectiveness of these methods on road networks \cite{alix2023pathletrl, tang2023explainable, chen2013pathlet}, our paper, to the best of our knowledge, is the first to adapt this approach within a context of content consumption on streaming platforms. 

\section{Background and Motivation}

In this section, we provide an overview of the key concepts and techniques that underpin our research, starting with dictionary learning and its extension to pathlet learning. We then discuss how pathlet learning, originally developed for geographical contexts, can be adapted to model user trajectories in music streaming data, with a focus on capturing the evolution of genre preferences over time. 

\label{section2}

\subsection{Dictionary and Pathlet Learning}
Dictionary learning (DL) is a representation learning technique that aims to learn a sparse representation of data. It has proven highly effective on various natural signals, including time-series \cite{marey2024modeling}, audio \cite{grosse2012shift}, and images \cite{mairal2009online, wohlberg2014efficient}. DL assumes that such signals are inherently sparse and can be expressed as a linear combination of a few recurring patterns, called \textit{atoms}.
These atoms represent basic and understandable elements or motifs in the data and are stored in a matrix known as the \textit{dictionary}.

Formally, given a set of $n$ observations $\mathbf{X} = [\mathbf{x}_1, \ldots, \mathbf{x}_n] \in \mathbb{R}^{n \times d}$, the objective of DL is to learn patterns by solving: \begin{equation}
\label{eq:DL} 
\mathcal{L}^*(\mathbf{X},D)=\min_{D \in \mathbb{R}^{d \times k}, \alpha \in \mathbb{R}^{k \times n}}\frac{1}{2}\Vert \mathbf{X}-D\alpha\Vert_F^2+\lambda \phi(\alpha)
\end{equation} where $D$ is the dictionary, $\alpha$ are the sparse coefficients, and $k \ll d$. The first term is the reconstruction error, and the second term enforces sparsity, typically through $\ell_0$ or $\ell_1$ norms for $\phi$. Here, $\|\cdot\|_F$ denotes the Frobenius norm, and $\lambda$ controls the trade-off between data fidelity and sparsity. 
This problem is typically solved using an alternating minimization approach, where we iteratively update the dictionary $D$ and the sparse coefficients $\alpha$ while keeping the other fixed, until convergence \cite{Lee2007}.

DL’s flexibility and transparency make it valuable for applications such as data compression \cite{skretting2011image}, denoising \cite{elad2006image}, anomaly detection \cite{pilastre2020anomaly}, and clustering \cite{sprechmann2010dictionary}. 
While DL is effective for structured data like images and time-series, modeling trajectories on non-Euclidean data (like paths on a graph) requires a more specialized approach. 
This is where Pathlet Learning (PL) comes into play.

PL was developed in geographical contexts, modeling trajectories on road networks, typically to ease traffic flow. 
Formally, given a graph $\mathcal{G}$, the set of observations is a collection of valid sequences along edges, like cab rides. 
The goal is to learn a dictionary $D$ of pathlets (recurring small paths on $\mathcal{G}$, similar to atoms in DL), helping reconstruct the sequences. 
In the literature, PL is framed as a constrained optimization problem:
in \cite{chen2013pathlet}, the authors used dynamic programming to satisfy constraints, while in \cite{tang2023explainable}, the problem is relaxed for projected gradient descent optimization. 
In \cite{kiermeier2017anomaly}, the authors addressed memory space complexity using reinforcement learning.

While previous research has focused on applications like data compression \cite{alix2023pathletrl, chen2013pathlet, tang2023explainable}, route planning \cite{chen2013pathlet, tang2023explainable}, and time journey prediction \cite{tang2023explainable}, we extend PL to model user musical taste trajectories, in order to capture how musical genre preferences evolve over time. 
Next, we discuss key considerations in applying PL to this new domain.

\subsection{From a Geographical Context to User Trajectories}
\label{stakes}

The paradigm of DL and PL is to represent data using a basis of atoms, defined as explicit and recurrent patterns within the dataset. 
In this sense, these approaches are inherently linked to the issue of transparency. 
Indeed, unlike black-box data representation methods, each coordinate of the embedding produced by DL or PL directly corresponds to an understandable atom.
Therefore, we aim to use PL to model how users engage with musical genres over time, highlighting typical listening patterns.

Reducing data dimensions through a small base of pathlets, PL can be seen as a data compression technique. 
Thus, high data sparsity increases the difficulty of data reconstruction.
In a geographical context, the graph is fixed, and the trajectories are valid paths on the graph. 
In a way, the nodes' semantics are strong, as very distant nodes cannot be connected. 
In contrast, in the context of music, nodes can be musical content, and user listening trajectories are built along the edges between these nodes. 
Unlike geographical nodes, which are semantically constrained by physical distance, musical content have weaker semantics, as two distant tracks (such as a metal track and a jazz one) can still be linked within a user's listening trajectory. 
This flexibility introduces additional challenges in modeling the transitions between genres, as the definition of the trajectories has a direct impact on the data sparsity. 
Furthermore, music streaming data exhibits long-tail distributions \cite{celma2009music, kowald2020unfairness}, where a few popular tracks accumulate the majority of streams, while a vast number of less-known tracks receive significantly fewer listens. 
For instance, in our dataset, $20 \%$ of the most popular songs account for more than $90 \%$ of the streams.
As a result, trajectories issued from music streaming data tend to be more sparse and relatively more difficult to reconstruct.

Our main problem can now be framed as follows:

\textbf{RQ: How can pathlet learning help capture user preferences in music context?}

\section{Pathlet Learning for User Trajectories}
\label{section3}
\label{PLalgo}
In this section, we extend the formalism of pathlet learning, shifting from a spatial to music content graphs.
\subsection{Problem Framing and Notations}
 
When it comes to music consumption, users are identified by a unique ID, and have their interactions with the content catalog tracked, along with a timestamp for each interaction. 
Formally, we consider a set of users $U$, a set of items (here, musical tracks) $I$, a set of musical genres $G$, and a time interval $T$ over which tracks were listened to by users. 

For each user $u$, we denote by $h_u$ their time-ordered listening history, containing triplets $(u, t, g) \in I \times T \times G$, where the user $u$ listened to track $i$ of genre $g$ at time $t$.

Furthermore, we slice $T$ in a number $K$ of fixed-size time windows of duration $\frac{|T|}{K}$.
In particular, we split the histories by K distinct time windows: $\forall k\in \{1, ..., K\}, \forall u \in U$,
\begin{equation*}
        h^{k}_u =  \Big\{ (i, t, g) \in h_u \Big|  \frac{k-1}{K}|T| \leq t <  \frac{k}{K}|T|\Big\}
\end{equation*}

We then define the \emph{genre allocation tensor} $X$ which, encodes users' taste towards genres over time windows, for $u \in U, g \in G, k \in \{1, ..., K\}$:

\begin{equation}
    X^{k}_{u, g} = \frac{\big|\{(i, t, g') \in h^{k}_u | g' = g \}\big|}{|h^{k}_u|}
    %Y_{u, j} = \frac{|\{(i, t') \in h_u | tag(i) = j, t' \in T_{t_{max}} \}|}{|\{(i, t') \in h_u | t' \in T_{t_{max}} \}|}.
\end{equation}

In other words, $X^{k}_{u, g}$ is the proportion of genre $g$ tracks listened to by user $u$ during the $k^{th}$ time window.
In the following, $X^{prev}$ and $Y$ stands for the allocation tensor during the penultimate and the last window, respectively.
Formally, we have:

\begin{equation*}
    X^{prev} = X^{K-1} \hspace{.4cm} Y = X^{K}
\end{equation*}

\begin{figure}[h!]
\centering
\includegraphics[width=0.9\linewidth]{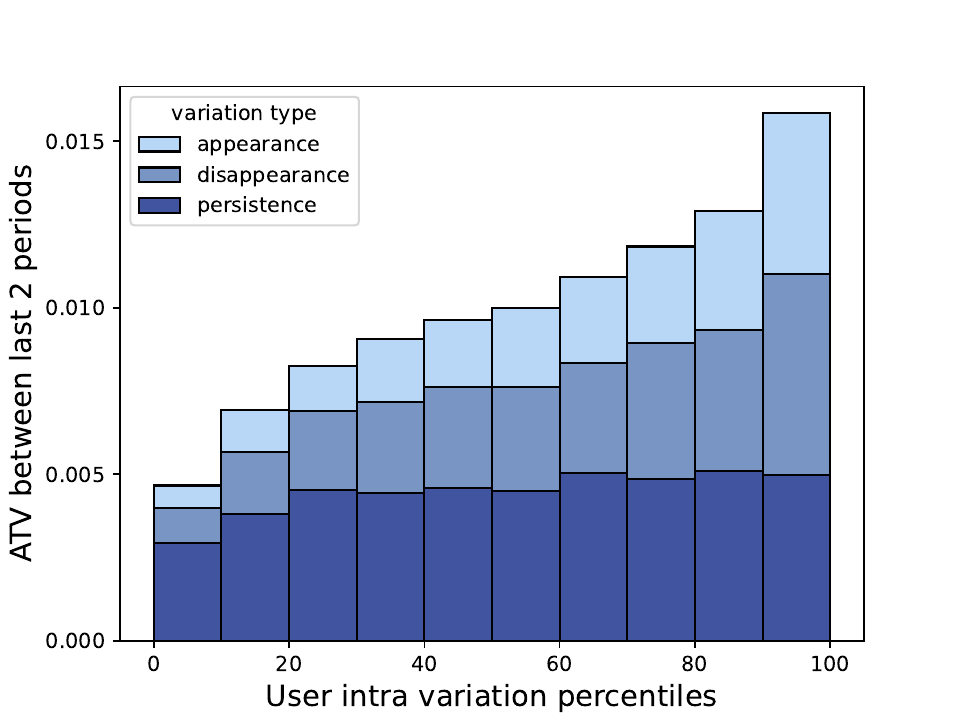}
\caption{Users' genre listening variation types between 2 successive periods with respect to intra-variability.
While \textit{appearance} and \textit{disappearance} correspond to emergence or abandonment of genres, persistence rely on variations of listening rate for a genre listened during 2 successive periods.
}
\label{fig:variation_types}
\end{figure}

In Figure~\ref{fig:variation_types}, we show how user genre allocation can evolve through 2 successive time windows, thanks to Average Total Variation (ATV, defined in Equation~\ref{eq:tv}).
The figure indicates that a significant share of variations is due to genre appearance and disappearance in our dataset. 
Being able to model the emergence and abandonment of genres in user histories is therefore of particular interest, both from a recommendation perspective and a sociological one, as mentioned in Section~\ref{intro}.
Our objective is thus to study the behavior of users with a focus on two phenomena: understanding how new genres appear in users' listening habits, and how they disappear.
Next, we define user trajectories accordingly. 

\subsection{Defining User Trajectories}

Here, our goal is to \textbf{predict next period taste vectors by revealing interactions between musical genres}.
We thus analyze the appearance and disappearance of such genres basing our trajectories on genre interaction, with the help of co-listening history vectors, defined for $u \in U, g \in G, k \in \{1, ...,K-1\}$ as follows: 

\begin{equation}
    \eta^{k}_{u, g} = \Big[\big|\{ (i_j, t_j, g_j) \in h^{k}_u  | (g', g) \in (g_j, g_{j \pm 1}) \}\big|\Big]_{g'\in G}
\end{equation}

In simpler terms, co-listening histories capture how often other musical genres are listened to immediately before or after the given genre $g$, effectively counting how genres `accompany' $g$ within the users' listening sequences.

From this history, and for a given pair $(u, g) \in U \times G$, we construct a list of genre positions over the time windows, leading to a $K-1$ sized discrete trajectory (the last window being the target). 
For each window $k$, a genre is sampled from $\eta^{k}_{u, g}$ co-listening counts, or, if no co-listened genre is available, from genre allocation values from $X^{k}_{u, g}$. 
For example, let us consider a user $u$ and $g = \textit{hard rock}$. 
We aim to trace a trajectory showing which genres interacted with hard rock for this user over time. 
For each time window, we randomly sample from the genres the user listened to either before or after a hard rock song. 
If the user did not listen to any hard rock songs during that window, we rely on its regular listening data to complete the information.

\paragraph{Building Trajectories}
As mentioned in section~\ref{stakes}, reducing sparsity is a key consideration before applying PL on user trajectories. For each user-genre pair $(u,g)$, we rank the user's most listened genres and place genre $g$ in position $0$, as it is the genre of interest. 
Next, we represent each trajectory as a sequence of ranks corresponding to the musical genres within it. 
Returning to our earlier example, suppose hard rock is listened to alongside metal at the beginning, with rock in the middle stages, and finally, around hard rock itself during the later stages.
The trajectory can then be formulated as follows:
\begin{center}
    $p = (metal, \dots, rock, \dots, \textit{hard rock})$
\end{center}
Assume that metal is the top genre for user $u$, with rock being the second most listened. 
Since hard rock is the genre of interest, the trajectory is transformed as $(1, \dots, 2, \dots, 0)$. 
This method helps to reduce sparsity, and suggests that musical genres can assume interchangeable roles in the emergence or decline of other genres, depending on their significance to the user.

\paragraph{Candidate Sets} We define the trajectories in such a way as to take into account which of the two objectives aforementioned is to be achieved. To build a trajectory set $P$, we repeatedly apply this trajectory sampling strategy over specific sets of pairs in $U \times G$.
In this case, we denote by 
\begin{equation*}
    A^+=\{(u, g) \in U \times G | X_{u,g}^{prev}=0 \land \exists k < K, X^{k}_{u,g} > 0\}
\end{equation*}
 the pairs $(u, g)$ such that $u$ already listened to $g$ at least once, but not during the penultimate period, making the $g$ in position of potential appearance at next period for the user. 
Similarly, we denote the pairs of candidates for disappearance by: 
\begin{equation*}
    A^{-}=\{(u, g) \in U \times G |X^{prev}_{u,g}>0\}
\end{equation*}
These are the user-genre pairs $(u, g)$ such that $u$ listened to $g$ during the penultimate period, placing $g$ as potential disappearing genre for $u$ in the next period.

\subsection{Learning the Dictionary}
%\todo[inline]{tiph : relire pour ajouter intuitions}
Next, depending on the event to be modeled (appearance or disappearance), a dictionary is computed from the trajectories issued for specific subsets of $U \times G$ pairs (namely $A^+$ and $A^-$). 
We refer to $P$ as the set of rank trajectories used for both cases, as the methodology remains the same.

\paragraph{Defining Graph and Paths' Representations} 
Let us define $\mathcal{G}$ as the graph induced by the trajectories of $P$. 
Formally, $\mathcal{G}=(E, V)$ with:

\begin{center}
        $E = \{ i \in \mathbbm{N} | \exists p \in P, i \in p \}$ \\
        \vspace{.15cm}
    $V = \{ (i, i') \in \mathbbm{N}^2| \exists p \in P, (i, i') \subset p \}$
\end{center}

\medskip
Specifically, $\mathcal{G}$ is the smallest directed graph such that any path in $P$ is valid in $\mathcal{G}$.
In addition, we define the set of candidate pathlets $D^{0}$ by computing recurrent sub-trajectories of $P$, \emph{i.e.} sub-paths with potential high reconstruction power. 
It is noteworthy that similar to the trajectories in $P$, candidate pathlets in $D^{0}$ are valid paths in $\mathcal{G}$.
Since paths on graphs are non-euclidean objects, performing optimization within a dictionary learning approach requires translating the various paths into an appropriate space, in this case through the one-hot encoding of paths along $\mathcal{G}$ edges.  
To this end, we define $\mathbf{P}$, the binary \textit{edge $\rightarrow$ paths} encoding matrix such that $\mathbf{P}_{j, j'} = \mathbbm{1}_{E_j \subset P_{j'}}$, and $\mathbf{D^{0}}$, the binary \textit{edge $\rightarrow$ candidate pathlets} matrix such that $\mathbf{D^{0}}_{j, j'} = \mathbbm{1}_{E_j \subset D^{0}_{j'}}$. 

\paragraph{Computing the Dictionary}
We aim to solve the following optimization problem:

\begin{equation}
\label{eq:PL}
    \mathcal{L}^*(\mathbf{P}, \mathbf{D^{0}})=\min_{\alpha \in [0, 1]^{|D^{0}| \times |P|}} \frac{1}{2}\Vert \mathbf{P}-\mathbf{D^{0}}\alpha \Vert^2_F +\lambda \Vert \alpha \Vert_1
\end{equation}
where $\alpha$ is the \textit{candidate pathlets $\rightarrow$ paths} encoding matrix and $\lambda \in \mathbb{R}$ plays the same role as in Equation~\ref{eq:DL}. \\

All elements of our optimization problem are sub-differentiable, so we minimize the loss using gradient descent, clipping $\alpha$ values to the $[0, 1]$ range at each step. 
The final dictionary $D$ is formed by selecting the top-$n$ pathlets most influential in reconstructing the trajectories, based on the $\alpha$ matrix. 
This allows flexible post hoc adjustments by retaining only the most relevant pathlets.

\textbf{Remark 1:} The condition $\mathbf{P} = \mathbf{D^{0}}\alpha$ guarantees full reconstruction of trajectories through the set of candidate pathlets. 
Thus, minimizing the reconstruction loss $\frac{1}{2} \Vert\mathbf{P} - \mathbf{D^{0}}\alpha\Vert^2_F$ enables us to capture much of the information contained in $P$ trajectories through pathlet encoding. 
The $\ell_1$ norm regularization promotes sparsity in the solution, emphasizing key pathlets for reconstruction.

\textbf{Remark 2:} In contrast to \cite{tang2023explainable}, we omit the additional loss term designed to reduce the number of pathlets in the encoding. While the authors argue that this term controls dictionary size, our experiments showed it weakly influences the optimization process and lacks clear evidence of an effect distinct from the sparsity loss. By excluding it, we simplify our model, reducing both parameters and computation time, while sticking more closely to traditional dictionary learning methods.

\subsection{Building Trajectory Embeddings}
In dictionary learning, the $\alpha$ matrix embeds input data ($P$ in our case).
But in this context, each coordinate of the $\alpha$ matrix corresponds to a candidate pathlet, which may not be part of the final dictionary.
To refine this, we can remove rows linked to unselected pathlets, forming a reduced matrix $\Tilde{\alpha}$, where columns represent trajectory embeddings. However, for a dictionary containing pathlets like $(rock, rap, jazz)$ and $(rock, jazz)$, the $\Tilde{\alpha}$ embedding for trajectory $(pop, rock, rap, jazz, pop)$ could activate both pathlets. This redundancy encodes the same information twice, reducing sparsity in the embeddings.
To overcome this, we separate the learning of dictionary from the embedding computation. 
Our trajectory embedding algorithm iteratively identifies the longest matching pathlet and updates the corresponding coordinate. By recursively segmenting the trajectory, we reduce redundancy, and prioritizing longer pathlets improves sparsity and expressiveness. This approach captures the sequence of the trajectory in explicit embeddings, useful for downstream tasks, as Figure~\ref{fig1} illustrates.

\section{Evolution of Musical Taste: Experimental Protocol}
\label{section4}

In this section, we describe the experimental protocol employed to evaluate our approach for modeling the evolution of musical taste. We first outline the evaluation metrics used to assess the relevance of the models, followed by the baselines chosen for comparison. Finally, we detail our proposed method and its specific implementation.

\subsection{Evaluation Metrics}
By using trajectory embeddings to predict last time window genre allocation $Y$, we check that the learned pathlets are relevant to model musical taste evolution. 
Note that this particular task was previously defined and studied in \cite{sanna2021next}. 
To compare the ground-truth last allocation $Y$ with the predicted one ($\hat{Y}$), the authors suggest using the average total variation, defined as follows:

\begin{equation}
\forall Y, \hat{Y}, \quad ATV(Y, \hat{Y}) = \frac{1}{n}\sum_{u\in U}\frac{1}{2}\sum_{g \in G}|Y_{u, g} - \hat{Y}_{u,g}|
\label{eq:tv}
\end{equation}
with $n$ being the number of users. In essence, $ATV$ measures the genre allocation absolute error, averaged on all users.
We also incorporate additional metrics derived from two binary classification subtasks presented in \cite{sanna2021next}.
Specifically:
\begin{itemize}
    \item \textit{\textbf{Plus-minus}} evaluates the model's ability to predict the direction of change in allocation, \emph{i.e.}, whether the proportion of a given genre increases or decreases.
    \item \textit{\textbf{New classes}} assesses the model's ability to predict the emergence of a new genre
\end{itemize}
These metrics use classical AUC score to evaluate models performances on these two subtasks.

\subsection{Baselines}
In \cite{sanna2021next}, authors compare their Preference Transition Model (\textsc{PTM}) with a naive strong baseline that consists in predicting the same $Y$ as the one observed in the penultimate time window ($X^{prev}$); this approach is referred to as \textsc{Previous} in the following. We also include the Non-negative Matrix Factorization (\textsc{NMF}, \cite{zhang2006learning}) which was the best candidate to compete with \textsc{PTM}. We add the {\textsc{Popularity}} baseline, which predicts for all users the same allocation, computed from the frequency of genres observed in training data. We implement all baselines from scratch, except for the \textsc{NMF}\footnote{\href{https://scikit-learn.org/stable/modules/generated/sklearn.decomposition.NMF.html}{NMF open source implementation}}. For \textsc{PTM}, despite our efforts to reproduce their results, the baseline performed unexpectedly poorly. The lack of released code further difficult reproduction, leading us to omit its scores.

\paragraph{Our Approach (\textsc{Plug-Previous})} 
On the one hand, \textsc{Previous} has proven effective for this task, consistently achieving the second-highest ATV scores in \cite{sanna2021next} without any learning. However, it is unable to predict any listening evolution. In contrast, our trajectory embeddings are specifically trained to discover the appearance or disappearance of musical genres. We thus start by predicting $\hat{Y}_{u,g}$ with \textsc{Previous}, which we will enrich with two sub-models, one for each detecting appearance and disappearance of genres. 
For the former one, we represent each user-genre candidate pair $(u, g) \in A^+$ averaging the learned embeddings of trajectories originating from $(u, g)$. These embeddings are then passed to a classifier to predict if the genre appeared between two sessions. 
If the model outputs a probability higher than 0.5, we assign to $\hat{Y}_{u,g}$ a new value, defined as the average value user $u$ allocated to $g$ during their whole history. 
We follow the same process for the disappearance, and we assign $\hat{Y}_{u,g} = 0$ for predicted disappearing pairs $(u, g) \in A^{-}$.
At last, normalizing $\hat{Y}_u$ vectors allows us to compare our predictions to $Y$. This approach is referred to as \textsc{Plug-Previous} in the following.  

\subsection{Hyperparameters}

\begin{figure}[h]
\centering
\includegraphics[width=0.9\linewidth]{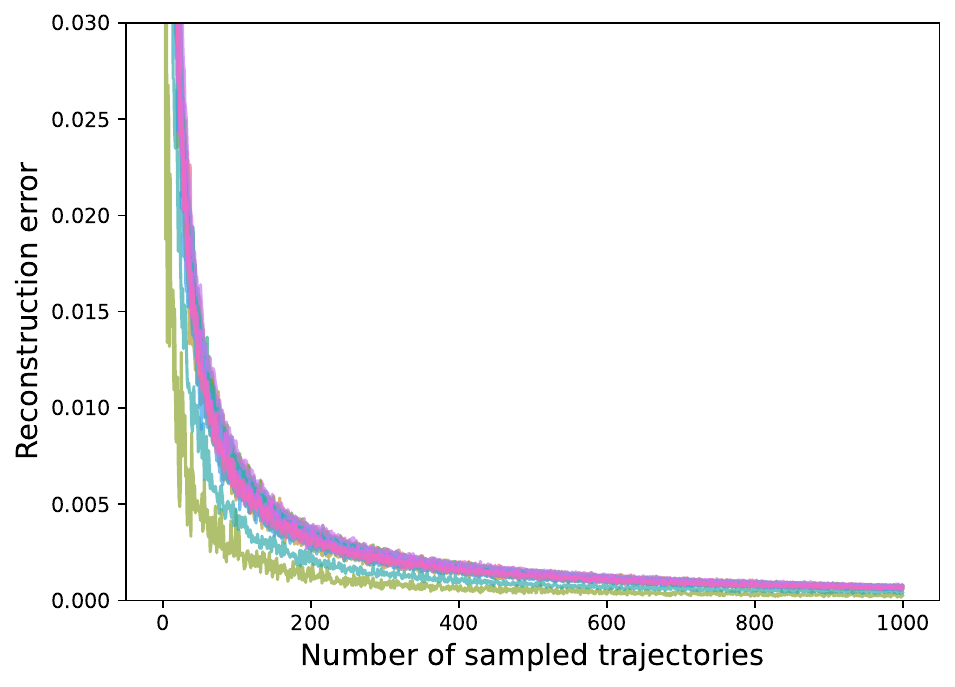}
\caption{
Reconstruction error of $X$ over the number of sampled trajectories.
}
\label{fig:sampling}
\end{figure}

Our hyperparameter choices are influenced by the constraints of available memory space and the necessity for interpretability in pathlet representations. 
Figure~\ref{fig:sampling} shows that sampling $1k$ trajectories per candidate from $A^+$ and $A^-$ capture a significant amount of co-listening trends.
For each of the two dictionaries, we randomly select $5k$ associated candidate trajectories. 
The candidate pathlets consist of the 10,000 most popular sub-paths of size less than 10 among the selected trajectories, allowing for a more manageable representation of user behavior.

\subsection{Datasets}

We apply our method on Deezer dataset, which contains $24M$ interactions between $2,000$ users and $350$ music genres over $17$ one-month time windows from January 2022 to June 2023. To expand our analysis, we apply our methodology to Last-fm\footnote{\href{http://ocelma.net/MusicRecommendationDataset/lastfm-1K.html}{http://ocelma.net/MusicRecommendationDataset/lastfm-1K.html}} dataset, containing $91,000$ interactions between $200$ users and $1,326$ music genres over $5$ three-months time windows from August 2007 to January 2019. 
The streams recorded in these datasets originate from organic listening and recommendation, as both are considered to provide meaningful insights into users' musical preferences.

\begin{figure}[h]
\centering
\includegraphics[width=0.9\linewidth]{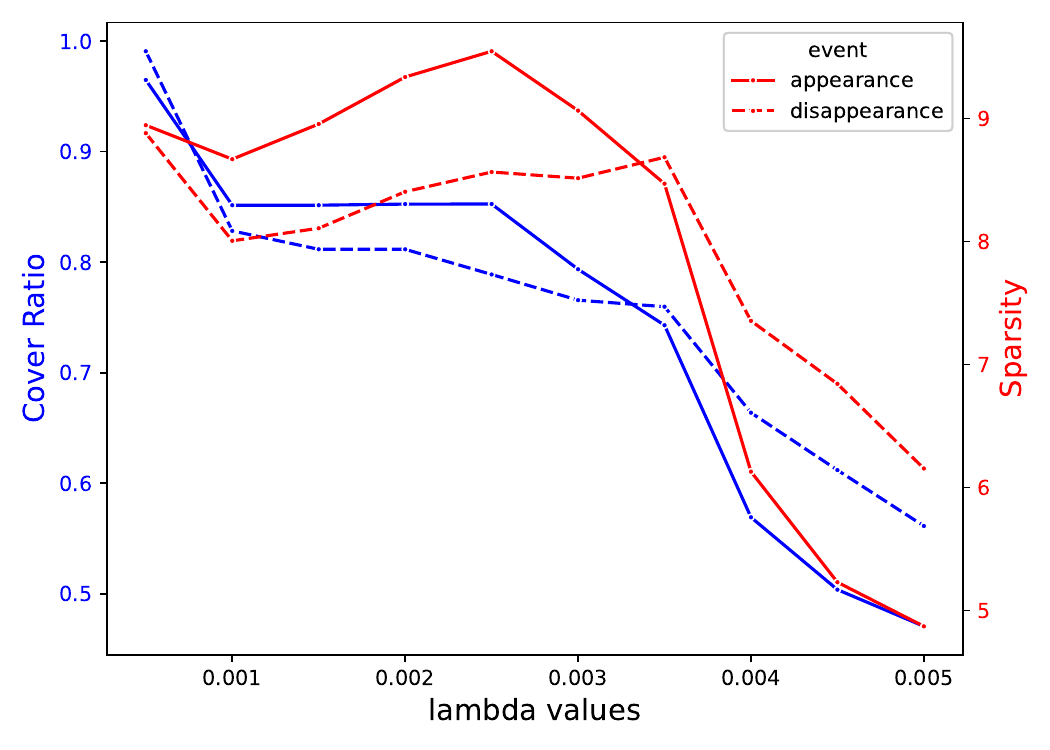}
\caption{
Sensitive Analysis: Impact of $\lambda$ on dictionary performances.
}
\label{fig:PL_results}
\end{figure}

Dictionaries are evaluated on 5k non-selected trajectories using classical PL metrics, specifically cover ratio \cite{tang2023explainable} and code sparsity. 
The sensitivity analysis prenseted in Figure~\ref{fig:PL_results} determined that a value of $\lambda=0.0025$  balances high reconstruction scores with embedding expressiveness in our PL model. 
We optimize the function presented in Equation~\ref{eq:PL} using the Adam optimizer with a learning rate of 0.01 and apply an early stopping criterion based on loss stagnation over 5 epochs. For the appearance and disappearance classification tasks, we utilize random forest classifiers, training them on the first $K-1$ time windows and assessing predictive performance on the final window.

\begin{figure*}[ht!]
\centering
\subfloat[\textit{hard rock} appearance induced graphs]{\includegraphics[width=0.35\linewidth]{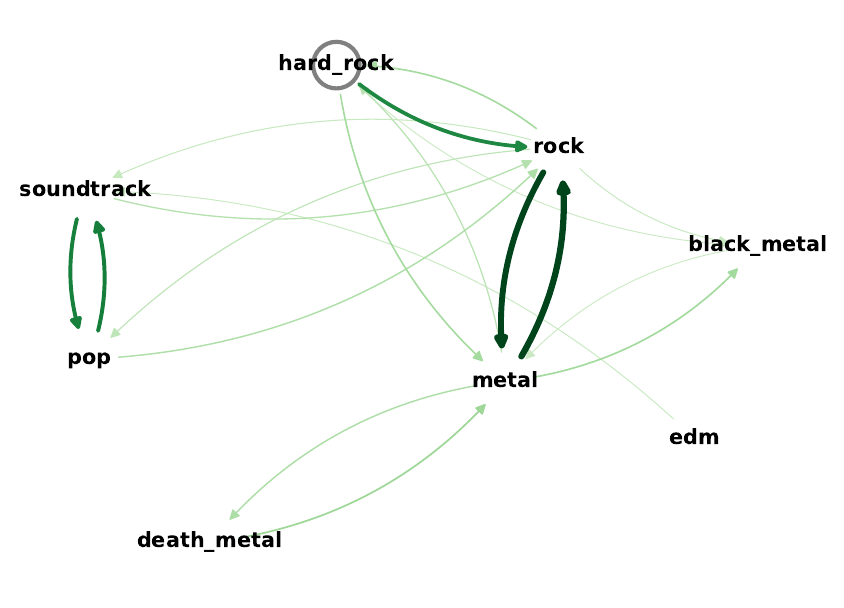}\label{fig:hard_rock_a}}
\subfloat[\textit{hard rock} disappearance induced graphs]{\includegraphics[width=0.35\linewidth]{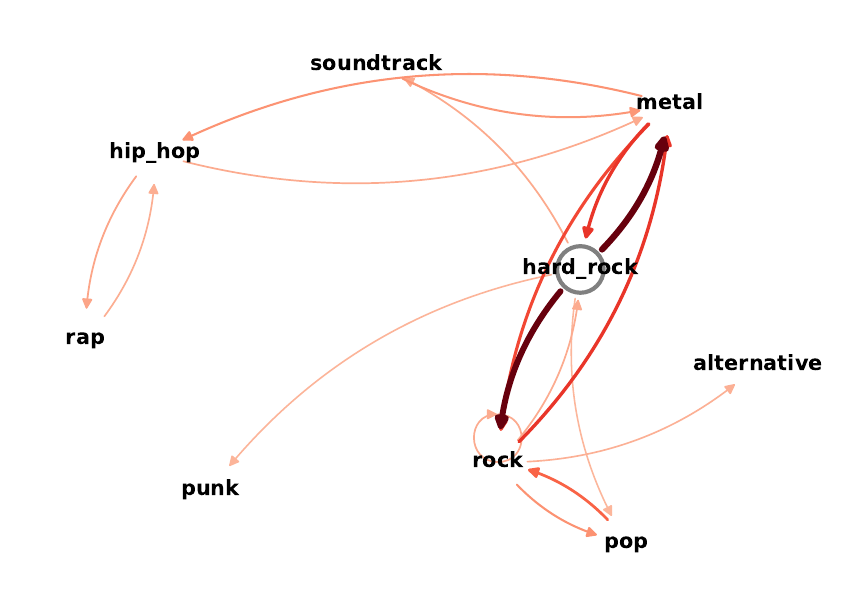}\label{fig:hard_rock_b}}\\
\subfloat[Diversity score of positive pathlets over genre popularity for appearance model]{\includegraphics[width=00.9\linewidth]{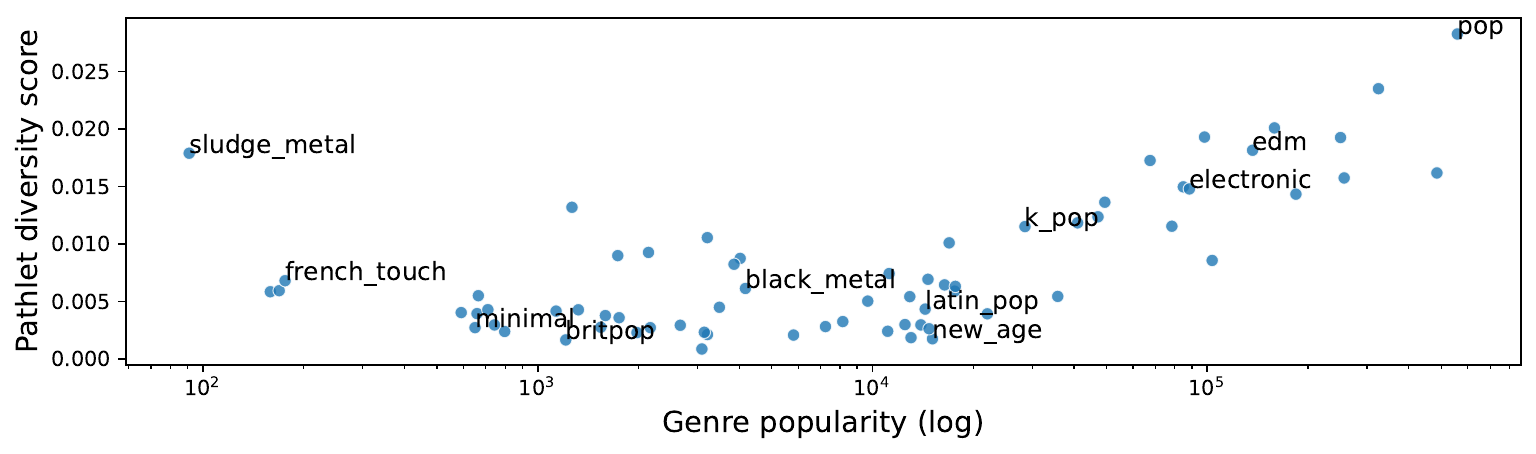}\label{fig:diversity}}
\caption{Pathlet interpretations : interactions between genres (top) and pathlet diversity (bottom)}
\end{figure*}

\section{Study on the Evolution of Musical Taste: Results}
\label{section5}

In this section, we present the results of our study on the evolution of musical taste, beginning with a quantitative evaluation of our model's performance on the task at hand. We then provide qualitative insights into genre dynamics, showing how pathlet embeddings help to analyze musical preferences over time.

\subsection{Quantitative Model Evaluation}

In Table~\ref{tab:pred_table}, the results of our \textsc{Previous-plug} prove that the introduction of the two models $appearance$ and $disappearance$ improves \textsc{Previous} baseline, reducing ATV for our dataset by acting on musical genres that are likely to appear or disappear.
We notice a significant increase for the \textbf{New classes} metric for both datasets, illustrating the performance of our $appearance$ prediction model. 
These results demonstrate that pathlet embeddings effectively capture relevant information regarding the evolution of genre allocations over time. 
Therefore, they justify the use of the computed embeddings for analyzing interactions between users and musical genres in the subsequent study.

\begin{table}[ht!]
\centering
\resizebox{0.5\textwidth}{!}{
\begin{tabular}{lcccc}
\toprule
\textbf{Dataset} & \textbf{Model} & ATV $(\downarrow)$ & Plus-minus $(\uparrow)$ & New classes $(\uparrow)$ \\
\midrule
\multirow{3}{*}{\begin{tabular}[c]{@{}c@{}}Deezer\end{tabular}} 
 & \textsc{Popularity} & $0.688$ & $0.840$ & $0.503$ \\
 & \textsc{NMF} & $0.400$ & $0.874$ & $0.503$ \\
 & \textsc{Previous} & $0.366$ & $0.900$* & $0.660$* \\
 & \textsc{Plug-Previous} & $\textbf{0.348}$ & $\textbf{0.908}$ & $\textbf{0.827}$ \\
\midrule
\multirow{3}{*}{Last-fm} 
 & \textsc{Popularity} & $0.842$ & $0.872$ & $0.503$ \\
 & \textsc{NMF} & $0.591$ & $0.884$ & $0.503$ \\
 & \textsc{Previous} & $\textbf{0.491}$ & $\textbf{0.917}$* & $0.681$* \\
 & \textsc{Plug-Previous} & $0.506$ & $0.907$ & $\textbf{0.999}$ \\

\bottomrule
\end{tabular}}
\caption{Prediction scores with all metrics for both datasets. 
* means that the results are computed with $1$ time window shift for metric validity.}
\label{tab:pred_table}
\end{table}

\subsection{Qualitative Analysis on Deezer dataset}
Now, we leverage the expressiveness of our embeddings to gain insights into the information captured by our approach.

\paragraph{Raw Pathlets Analysis}
First, we study the correlations of the embedding dimensions (each of which refers to a pathlet) with the outputs of our two sub-tasks. We characterize a pathlet by the presence of the target genre, defining \textbf{inertial pathlets} (in those, rank $0$ genre is present, indicating that the genre has been self-accompanying over the user's history, showing certain inertia), and by the average rank of the genres present in the pathlet, defining \textbf{listening regime} (as high ranks refer to low listening rates by the user, the pathlet average rank indicates if the target genre was listened to within a minority listening regime for the user or not). 

\begin{figure}[t!]
\centering
\includegraphics[width=\linewidth]{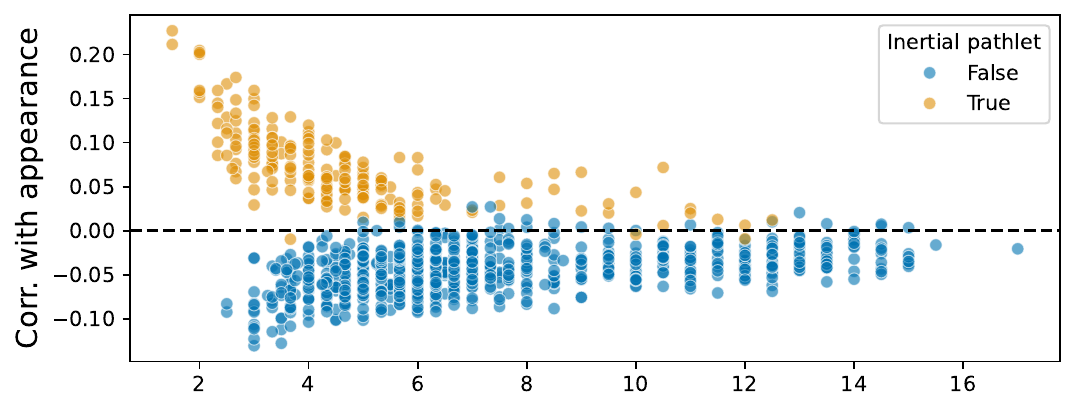}
\includegraphics[width=\linewidth]{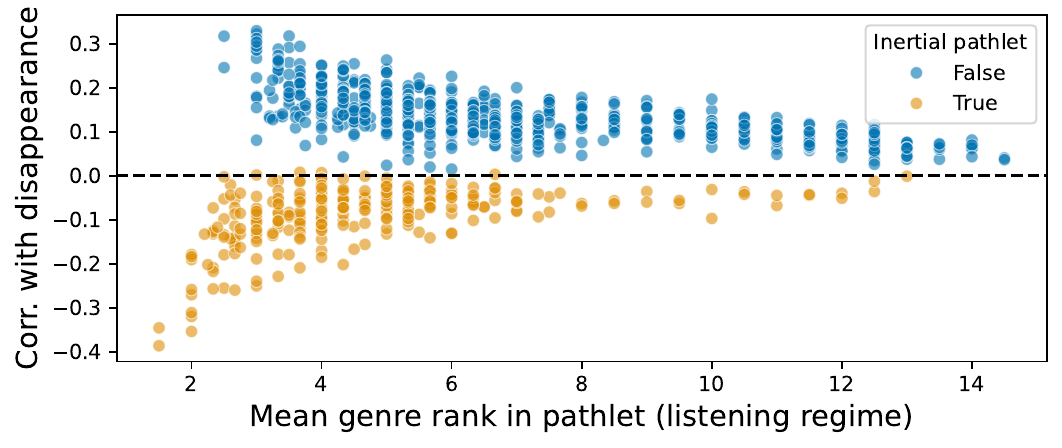} 
\caption{Pathlets correlation with appearance (top) and disappearance (bottom).
}
\label{fig:pathlets_analysis}
\end{figure}

In Figure~\ref{fig:pathlets_analysis}, the pathlet correlations show symmetrical insights for both appearance and disappearance.
Inertial pathlets positively correlate with retention (appearance or non-disappearance), whereas non-inertial pathlets exhibit a negative correlation. In other words, for the same number of listens within a given period, a genre with consecutive listens is more likely to remain than one where the listens are scattered throughout the period. Furthermore, the more a pathlet embeds a minority listening regime, the more polarized its influence. This means that the correlation with retention increases as an inertial pathlet embeds a minority listening regime, but decreases for a non-inertial pathlet under the same condition.

\paragraph{Extended Pathlets Analysis}
Next, we invert the ranking operation from trajectory sampling to reveal extended pathlets, enabling genre-based analysis. 
By weighting extended pathlets based on their values in embedding components, and on their correlations with appearance and disappearance, we generate weighted graphs that highlight genre dynamics. For example, in the case of \textit{hard rock} (Figure~\ref{fig:hard_rock_a}), we observe that co-listening to \textit{rock} and \textit{metal} has a significant impact on its retention.

Paths connecting \textit{hard rock} to \textit{rock} and having strong links with \textit{metal} lead to the appearance of \textit{hard rock} (Figure~\ref{fig:hard_rock_a}). Conversely, paths from \textit{hard rock} to \textit{rock} or \textit{metal} with weaker connections between \textit{rock} and \textit{metal} can cause \textit{hard rock} to disappear (Figure~\ref{fig:hard_rock_b}). This suggests that \textit{hard rock} competes with both \textit{rock} and \textit{metal} individually, but listening to these two genres together reinforces \textit{hard rock}'s presence in listening histories.

We can study the characteristics of extended pathlets, \emph{e.g.}, by developing a diversity score that quantifies the extent to which certain pathlets contain several distinct genres.
This score can then be linked to the pathlets correlations with the emergence of new genres. 
This approach allows us to better understand the role of diversity in music listening, and how it shapes genre appearance dynamics.
In Figure~\ref{fig:diversity}, both popular and niche genres emerge from highly diversified trajectories. 
It shows that popular genres persist in listening patterns due to their universality, even when users diverge, while niche genres require deliberate exploration by curious users seeking less mainstream content.

\section{Conclusion and Future Work}

In this paper, we introduced a novel pathlet learning approach for modeling user trajectories on music platforms. By constructing pathlet dictionaries and utilizing trajectory embeddings, we predicted future genre allocations, demonstrating how pathlets can be understood and reveal meaningful patterns in user listening behavior. 
We believe this work paves the way for not only a better understanding of the dynamics of musical taste, but also for more interpretable music consumption models, encouraging research that prioritizes understandability. 
By providing clearer insights into user listening patterns and genre interactions, our approach has the potential to enhance recommender systems, offering users greater transparency and a better understanding of their personalized recommendations.
For example, pathlet learning could enhance playlist recommendation by expanding user taste and promoting diversity in suggested tracks. 
Another potential application is its integration into long-term taste modeling within next-item recommendation frameworks, helping balance short-term preferences with evolving listening habits. 
Moreover, in cold-start recommendation, the technique’s ability to capture recurring patterns could provide valuable prior knowledge for new users or less active listeners. 
Finally, diversity-aware recommendation could benefit from pathlet learning by ensuring that recommendations reflect evolving genre exploration patterns, guiding users along learned trajectories.

As future work, a natural extension of this research would be to apply the models to other domains, such as movies or TV shows. 
But as mentioned in Section~\ref{stakes}, pathlet learning is most effective in a context of dense data.
In a way, the music domain is well-suited for this method due to the repetitive nature of music consumption, which allows natural emergence of recurring patterns.
Also, this domain presents numerous interactions with many different items, which is not always the case in other domains.
For example, book or movie consumption have quite different patterns, with rather scarcer and less repetitive interaction data.
Switching to such a domain thus may present challenges in achieving the same reconstruction level, due to low level of interaction density between users and items.
Therefore, from an algorithmic perspective, a promising direction for future work would be to explore methods for addressing the sparsity inherent to other domains within the context of pathlet learning, for example developing specific dimension reduction techniques to better handle sparse interaction matrices.
Finally, another track for future research would involve evaluating the visualization offered by the proposed framework as a tool for social scientists, such as musicologists, to investigate consumption patterns. 

\section*{Acknowledgements}
This paper has been partially realized in the framework of the “RECORDS” grant (ANR-2019-CE38-0013) funded by the ANR (French National Agency of Research)

% \printbibliography
\bibliography{refs}

\begin{thebibliography}{53}
\providecommand{\natexlab}[1]{#1}
\providecommand{\url}[1]{\texttt{#1}}
\expandafter\ifx\csname urlstyle\endcsname\relax
  \providecommand{\doi}[1]{doi: #1}\else
  \providecommand{\doi}{doi: \begingroup \urlstyle{rm}\Url}\fi

\bibitem[Cura et~al.(2022)Cura, Beaumont, Beuscart, Coavoux, Latreille~de Fozi{\`e}res, Le~Bigot, Renisio, Moussallam, and Louail]{cura2022uplifting}
Robin Cura, Am{\'e}lie Beaumont, Jean-Samuel Beuscart, Samuel Coavoux, No{\'e} Latreille~de Fozi{\`e}res, Brenda Le~Bigot, Yann Renisio, Manuel Moussallam, and Thomas Louail.
\newblock Uplifting interviews in social science with individual data visualization: The case of music listening.
\newblock In \emph{CHI Conference on Human Factors in Computing Systems Extended Abstracts}, pages 1--9, 2022.

\bibitem[Moore et~al.(2013)Moore, Chen, Turnbull, and Joachims]{moore2013taste}
Joshua~L Moore, Shuo Chen, Douglas~R Turnbull, and Thorsten Joachims.
\newblock Taste over time: The temporal dynamics of user preferences.
\newblock In \emph{ISMIR}, volume~13, page 406, 2013.

\bibitem[Petridis et~al.(2022)Petridis, Daskalova, Mennicken, Way, Lamere, and Thom]{petridis2022tastepaths}
Savvas Petridis, Nediyana Daskalova, Sarah Mennicken, Samuel~F Way, Paul Lamere, and Jennifer Thom.
\newblock Tastepaths: Enabling deeper exploration and understanding of personal preferences in recommender systems.
\newblock In \emph{Proceedings of the 27th International Conference on Intelligent User Interfaces}, pages 120--133, 2022.

\bibitem[Seaver(2022)]{seaver2022computing}
Nick Seaver.
\newblock \emph{Computing taste: Algorithms and the makers of music recommendation}.
\newblock University of Chicago Press, 2022.

\bibitem[Villermet et~al.(2021)Villermet, Poiroux, Moussallam, Louail, and Roth]{villermet2021follow}
Quentin Villermet, J{\'e}r{\'e}mie Poiroux, Manuel Moussallam, Thomas Louail, and Camille Roth.
\newblock Follow the guides: disentangling human and algorithmic curation in online music consumption.
\newblock In \emph{Proceedings of the 15th ACM Conference on Recommender Systems}, pages 380--389, 2021.

\bibitem[Bourdieu(2018)]{bourdieu2018distinction}
Pierre Bourdieu.
\newblock Distinction a social critique of the judgement of taste.
\newblock In \emph{Inequality}, pages 287--318. Routledge, 2018.

\bibitem[DeNora(2000)]{denora2000music}
Tia DeNora.
\newblock \emph{Music in everyday life}.
\newblock Cambridge university press, 2000.

\bibitem[Hesmondhalgh(2013)]{hesmondhalgh2013music}
David Hesmondhalgh.
\newblock \emph{Why music matters}.
\newblock John Wiley \& Sons, 2013.

\bibitem[McCourt and Zuberi(2016)]{mccourt2016music}
Tom McCourt and Nabeel Zuberi.
\newblock Music and discovery.
\newblock \emph{Popular Communication}, 14\penalty0 (3):\penalty0 123--126, 2016.

\bibitem[Riom(2020)]{riom2020discovering}
Loïc Riom.
\newblock Discovering music at sofar sounds: Surprise, attachment, and the fan--artist relationship.
\newblock \emph{Popular Music, Technology, and the Changing Media Ecosystem: From Cassettes to Stream}, pages 201--216, 2020.

\bibitem[Schedl et~al.(2018)Schedl, Zamani, Chen, Deldjoo, and Elahi]{schedl2018current}
Markus Schedl, Hamed Zamani, Ching-Wei Chen, Yashar Deldjoo, and Mehdi Elahi.
\newblock Current challenges and visions in music recommender systems research.
\newblock \emph{International Journal of Multimedia Information Retrieval}, 7:\penalty0 95--116, 2018.

\bibitem[Bucher(2016)]{bucher2016neither}
Taina Bucher.
\newblock Neither black nor box: Ways of knowing algorithms.
\newblock \emph{Innovative methods in media and communication research}, pages 81--98, 2016.

\bibitem[Lesota et~al.(2022)Lesota, Parada-Cabaleiro, Brandl, Lex, Rekabsaz, and Schedl]{lesota2022traces}
Oleg Lesota, Emilia Parada-Cabaleiro, Stefan Brandl, Elisabeth Lex, Navid Rekabsaz, and Markus Schedl.
\newblock Traces of globalization in online music consumption patterns and results of recommendation algorithms.
\newblock In \emph{ISMIR}, pages 291--297, 2022.

\bibitem[Matrosova et~al.(2024)Matrosova, Marey, Salha-Galvan, Louail, Bodini, and Moussallam]{matrosova2024recommender}
Kristina Matrosova, Lilian Marey, Guillaume Salha-Galvan, Thomas Louail, Olivier Bodini, and Manuel Moussallam.
\newblock Do recommender systems promote local music? a reproducibility study using music streaming data.
\newblock In \emph{Proceedings of the 18th ACM Conference on Recommender Systems}, pages 148--157, 2024.

\bibitem[Ungruh et~al.(2024)Ungruh, Dinnissen, Volk, Pera, and Hauptmann]{ungruh2024putting}
Robin Ungruh, Karlijn Dinnissen, Anja Volk, Maria~Soledad Pera, and Hanna Hauptmann.
\newblock Putting popularity bias mitigation to the test: A user-centric evaluation in music recommenders.
\newblock In \emph{Proceedings of the 18th ACM Conference on Recommender Systems}, pages 169--178, 2024.

\bibitem[Vrijenhoek et~al.(2024)Vrijenhoek, Daniil, Sandel, and Hollink]{vrijenhoek2024diversity}
Sanne Vrijenhoek, Savvina Daniil, Jorden Sandel, and Laura Hollink.
\newblock Diversity of what? on the different conceptualizations of diversity in recommender systems.
\newblock In \emph{The 2024 ACM Conference on Fairness, Accountability, and Transparency}, pages 573--584, 2024.

\bibitem[Smith et~al.(2024)Smith, Satwani, Burke, and Fiesler]{smith2024recommend}
Jessie~J Smith, Aishwarya Satwani, Robin Burke, and Casey Fiesler.
\newblock Recommend me? designing fairness metrics with providers.
\newblock In \emph{The 2024 ACM Conference on Fairness, Accountability, and Transparency}, pages 2389--2399, 2024.

\bibitem[Ekstrand et~al.(2014)Ekstrand, Harper, Willemsen, and Konstan]{ekstrand2014user}
Michael~D Ekstrand, F~Maxwell Harper, Martijn~C Willemsen, and Joseph~A Konstan.
\newblock User perception of differences in recommender algorithms.
\newblock In \emph{Proceedings of the 8th ACM Conference on Recommender systems}, pages 161--168, 2014.

\bibitem[Afchar et~al.(2022)Afchar, Melchiorre, Schedl, Hennequin, Epure, and Moussallam]{afchar2022explainability}
Darius Afchar, Alessandro Melchiorre, Markus Schedl, Romain Hennequin, Elena Epure, and Manuel Moussallam.
\newblock Explainability in music recommender systems.
\newblock \emph{AI Magazine}, 43\penalty0 (2):\penalty0 190--208, 2022.

\bibitem[Zhou et~al.(2008)Zhou, Wilkinson, Schreiber, and Pan]{zhou2008large}
Yunhong Zhou, Dennis Wilkinson, Robert Schreiber, and Rong Pan.
\newblock Large-scale parallel collaborative filtering for the netflix prize.
\newblock In \emph{Algorithmic Aspects in Information and Management: 4th International Conference, AAIM 2008, Shanghai, China, June 23-25, 2008. Proceedings 4}, pages 337--348. Springer, 2008.

\bibitem[Hu et~al.(2008)Hu, Koren, and Volinsky]{hu2008collaborative}
Yifan Hu, Yehuda Koren, and Chris Volinsky.
\newblock Collaborative filtering for implicit feedback datasets.
\newblock In \emph{2008 Eighth IEEE international conference on data mining}, pages 263--272. Ieee, 2008.

\bibitem[Tak{\'a}cs and Tikk(2012)]{takacs2012alternating}
G{\'a}bor Tak{\'a}cs and Domonkos Tikk.
\newblock Alternating least squares for personalized ranking.
\newblock In \emph{Proceedings of the sixth ACM conference on Recommender systems}, pages 83--90, 2012.

\bibitem[Van~Meteren and Van~Someren(2000)]{van2000using}
Robin Van~Meteren and Maarten Van~Someren.
\newblock Using content-based filtering for recommendation.
\newblock In \emph{Proceedings of the machine learning in the new information age: MLnet/ECML2000 workshop}, volume~30, pages 47--56. Barcelona, 2000.

\bibitem[Burke(2002)]{burke2002hybrid}
Robin Burke.
\newblock Hybrid recommender systems: Survey and experiments.
\newblock \emph{User modeling and user-adapted interaction}, 12:\penalty0 331--370, 2002.

\bibitem[Jawaheer et~al.(2014)Jawaheer, Weller, and Kostkova]{jawaheer2014modeling}
Gawesh Jawaheer, Peter Weller, and Patty Kostkova.
\newblock Modeling user preferences in recommender systems: A classification framework for explicit and implicit user feedback.
\newblock \emph{ACM Transactions on Interactive Intelligent Systems (TiiS)}, 4\penalty0 (2):\penalty0 1--26, 2014.

\bibitem[He et~al.(2017)He, Liao, Zhang, Nie, Hu, and Chua]{he2017neural}
Xiangnan He, Lizi Liao, Hanwang Zhang, Liqiang Nie, Xia Hu, and Tat-Seng Chua.
\newblock Neural collaborative filtering.
\newblock In \emph{Proceedings of the 26th international conference on world wide web}, pages 173--182, 2017.

\bibitem[Medsker et~al.(2001)Medsker, Jain, et~al.]{medsker2001recurrent}
Larry~R Medsker, Lakhmi Jain, et~al.
\newblock Recurrent neural networks.
\newblock \emph{Design and Applications}, 5\penalty0 (64-67):\penalty0 2, 2001.

\bibitem[Graves and Graves(2012)]{graves2012long}
Alex Graves and Alex Graves.
\newblock Long short-term memory.
\newblock \emph{Supervised sequence labelling with recurrent neural networks}, pages 37--45, 2012.

\bibitem[Tran et~al.(2024)Tran, Salha-Galvan, Sguerra, and Hennequin]{tran2024transformers}
Viet-Anh Tran, Guillaume Salha-Galvan, Bruno Sguerra, and Romain Hennequin.
\newblock Transformers meet act-r: Repeat-aware and sequential listening session recommendation.
\newblock In \emph{Proceedings of the 18th ACM Conference on Recommender Systems}, pages 486--496, 2024.

\bibitem[Devooght and Bersini(2017)]{devooght2017long}
Robin Devooght and Hugues Bersini.
\newblock Long and short-term recommendations with recurrent neural networks.
\newblock In \emph{Proceedings of the 25th conference on user modeling, adaptation and personalization}, pages 13--21, 2017.

\bibitem[Deepak and Prasad(2020)]{deepak2020music}
S~Deepak and BG~Prasad.
\newblock Music classification based on genre using lstm.
\newblock In \emph{2020 Second International Conference on Inventive Research in Computing Applications (ICIRCA)}, pages 985--991. IEEE, 2020.

\bibitem[Romero and Bie(2022)]{RomeroB22}
Rapha{\"{e}}l Romero and Tijl~De Bie.
\newblock Embedding-based next song recommendation for playlists.
\newblock In \emph{European Symposium on Artificial Neural Networks, Computational Intelligence and Machine Learning}, 2022.

\bibitem[Sanna~Passino et~al.(2021)Sanna~Passino, Maystre, Moor, Anderson, and Lalmas]{sanna2021next}
Francesco Sanna~Passino, Lucas Maystre, Dmitrii Moor, Ashton Anderson, and Mounia Lalmas.
\newblock Where to next? a dynamic model of user preferences.
\newblock In \emph{Proceedings of the Web Conference 2021}, pages 3210--3220, 2021.

\bibitem[Zhang et~al.(2006)Zhang, Wang, Ford, and Makedon]{zhang2006learning}
Sheng Zhang, Weihong Wang, James Ford, and Fillia Makedon.
\newblock Learning from incomplete ratings using non-negative matrix factorization.
\newblock In \emph{Proceedings of the 2006 SIAM international conference on data mining}, pages 549--553. SIAM, 2006.

\bibitem[Gopalan et~al.(2015)Gopalan, Hofman, and Blei]{gopalan2015scalable}
Prem Gopalan, Jake~M Hofman, and David~M Blei.
\newblock Scalable recommendation with hierarchical poisson factorization.
\newblock In \emph{UAI}, pages 326--335, 2015.

\bibitem[Guo et~al.(2020)Guo, Zhuang, Qin, Zhu, Xie, Xiong, and He]{guo2020survey}
Qingyu Guo, Fuzhen Zhuang, Chuan Qin, Hengshu Zhu, Xing Xie, Hui Xiong, and Qing He.
\newblock A survey on knowledge graph-based recommender systems.
\newblock \emph{IEEE Transactions on Knowledge and Data Engineering}, 34\penalty0 (8):\penalty0 3549--3568, 2020.

\bibitem[Hou and Shi(2019)]{hou2019explainable}
Hao Hou and Chongyang Shi.
\newblock Explainable sequential recommendation using knowledge graphs.
\newblock In \emph{Proceedings of the 5th International Conference on Frontiers of Educational Technologies}, pages 53--57, 2019.

\bibitem[Wu et~al.(2023)Wu, Fang, Sun, Geng, Kong, and Ong]{wu2023generic}
Huizi Wu, Hui Fang, Zhu Sun, Cong Geng, Xinyu Kong, and Yew-Soon Ong.
\newblock A generic reinforced explainable framework with knowledge graph for session-based recommendation.
\newblock In \emph{2023 IEEE 39th International Conference on Data Engineering (ICDE)}, pages 1260--1272. IEEE, 2023.

\bibitem[Mairal et~al.(2009)Mairal, Bach, Ponce, and Sapiro]{mairal2009online}
Julien Mairal, Francis Bach, Jean Ponce, and Guillermo Sapiro.
\newblock Online dictionary learning for sparse coding.
\newblock In \emph{Proceedings of the 26th annual international conference on machine learning}, pages 689--696, 2009.

\bibitem[Wohlberg(2014)]{wohlberg2014efficient}
Brendt Wohlberg.
\newblock Efficient convolutional sparse coding.
\newblock In \emph{2014 IEEE International Conference on Acoustics, Speech and Signal Processing (ICASSP)}, pages 7173--7177. IEEE, 2014.

\bibitem[Grosse et~al.(2012)Grosse, Raina, Kwong, and Ng]{grosse2012shift}
Roger Grosse, Rajat Raina, Helen Kwong, and Andrew~Y Ng.
\newblock Shift-invariance sparse coding for audio classification.
\newblock \emph{arXiv preprint arXiv:1206.5241}, 2012.

\bibitem[Marey et~al.(2024)Marey, Sguerra, and Moussallam]{marey2024modeling}
Lilian Marey, Bruno Sguerra, and Manuel Moussallam.
\newblock Modeling activity-driven music listening with pace.
\newblock In \emph{Proceedings of the 2024 Conference on Human Information Interaction and Retrieval}, pages 346--351, 2024.

\bibitem[Alix and Papagelis(2023)]{alix2023pathletrl}
Gian Alix and Manos Papagelis.
\newblock Pathletrl: Trajectory pathlet dictionary construction using reinforcement learning.
\newblock In \emph{Proceedings of the 31st ACM International Conference on Advances in Geographic Information Systems}, pages 1--12, 2023.

\bibitem[Tang et~al.(2023)Tang, Peng, and Li]{tang2023explainable}
Yuanbo Tang, Zhiyuan Peng, and Yang Li.
\newblock Explainable trajectory representation through dictionary learning.
\newblock In \emph{Proceedings of the 31st ACM International Conference on Advances in Geographic Information Systems}, pages 1--4, 2023.

\bibitem[Chen et~al.(2013)Chen, Su, Huang, Zhang, and Guibas]{chen2013pathlet}
Chen Chen, Hao Su, Qixing Huang, Lin Zhang, and Leonidas Guibas.
\newblock Pathlet learning for compressing and planning trajectories.
\newblock In \emph{Proceedings of the 21st ACM SIGSPATIAL International Conference on Advances in Geographic Information Systems}, pages 392--395, 2013.

\bibitem[Lee et~al.(2006)Lee, Battle, Raina, and Ng]{Lee2007}
Honglak Lee, Alexis Battle, Rajat Raina, and Andrew~Y. Ng.
\newblock Efficient sparse coding algorithms.
\newblock In \emph{Proceedings of the 19th International Conference on Neural Information Processing Systems}, NIPS'06, page 801–808, Cambridge, MA, USA, 2006. MIT Press.

\bibitem[Skretting and Engan(2011)]{skretting2011image}
Karl Skretting and Kjersti Engan.
\newblock Image compression using learned dictionaries by rls-dla and compared with k-svd.
\newblock In \emph{2011 IEEE International Conference on Acoustics, Speech and Signal Processing (ICASSP)}, pages 1517--1520. IEEE, 2011.

\bibitem[Elad and Aharon(2006)]{elad2006image}
Michael Elad and Michal Aharon.
\newblock Image denoising via sparse and redundant representations over learned dictionaries.
\newblock \emph{IEEE Transactions on Image processing}, 15\penalty0 (12):\penalty0 3736--3745, 2006.

\bibitem[Pilastre et~al.(2020)Pilastre, Boussouf, d’Escrivan, and Tourneret]{pilastre2020anomaly}
Barbara Pilastre, Loic Boussouf, St{\'e}phane d’Escrivan, and Jean-Yves Tourneret.
\newblock Anomaly detection in mixed telemetry data using a sparse representation and dictionary learning.
\newblock \emph{Signal Processing}, 168:\penalty0 107320, 2020.

\bibitem[Sprechmann and Sapiro(2010)]{sprechmann2010dictionary}
Pablo Sprechmann and Guillermo Sapiro.
\newblock Dictionary learning and sparse coding for unsupervised clustering.
\newblock In \emph{2010 IEEE international conference on acoustics, speech and signal processing}, pages 2042--2045. IEEE, 2010.

\bibitem[Kiermeier et~al.(2017)Kiermeier, Werner, Linnhoff-Popien, Sauer, and Wieghardt]{kiermeier2017anomaly}
Marie Kiermeier, Martin Werner, Claudia Linnhoff-Popien, Horst Sauer, and Jan Wieghardt.
\newblock Anomaly detection in self-organizing industrial systems using pathlets.
\newblock In \emph{2017 IEEE International Conference on Industrial Technology (ICIT)}, pages 1226--1231. IEEE, 2017.

\bibitem[Celma~Herrada et~al.(2009)]{celma2009music}
{\`O}scar Celma~Herrada et~al.
\newblock \emph{Music recommendation and discovery in the long tail}.
\newblock Universitat Pompeu Fabra, 2009.

\bibitem[Kowald et~al.(2020)Kowald, Schedl, and Lex]{kowald2020unfairness}
Dominik Kowald, Markus Schedl, and Elisabeth Lex.
\newblock The unfairness of popularity bias in music recommendation: A reproducibility study.
\newblock In \emph{Advances in Information Retrieval: 42nd European Conference on IR Research, ECIR 2020, Lisbon, Portugal, April 14--17, 2020, Proceedings, Part II 42}, pages 35--42. Springer, 2020.

\end{thebibliography}
\end{document}